\documentclass[twocolumn,showpacs,preprintnumbers,amsmath,amssymb]{revtex4}

\usepackage{graphicx} 
\usepackage{dcolumn} 
\usepackage{bm} 

\begin{document}


\title{Thermoelectric effects in strongly interacting quantum dot coupled to 
       ferromagnetic leads}

\author{M. Krawiec}

 \email{krawiec@kft.umcs.lublin.pl}

\author{K. I. Wysoki\'{n}ski}

 \email{karol@tytan.umcs.lublin.pl}

\affiliation{Institute of Physics and Nanotechnology Center, 
             M. Curie-Sk\l{}odowska University, Pl. M. Curie-Sk\l odowskiej 1,
	     20-031 Lublin, Poland}

\date{\today}

\begin{abstract}
We study thermoelectric effects in Kondo correlated quantum dot coupled to 
ferromagnetic electrodes by calculating conductance, thermopower and thermal 
conductance in the Kondo regime. We also study the effect of the asymmetry in 
the coupling to the leads, which has important consequences for anti-parallel 
magnetization configuration. We discuss the thermoelectric figure of merit, 
tunnel magnetoresistance and violation of the Wiedemann-Franz law in this 
system. The results agree with recently measured thermopower of the quantum
dot defined in a two dimensional electron gas. 
\end{abstract}
\pacs{75.20.Hr, 72.15.Qm, 72.25.-b, 73.23.Hk}

\maketitle

\section{Introduction}
Transport properties of the device consisting of the quantum dot attached to 
external leads are strongly affected by the appearance of the correlated many 
body Kondo state \cite{Phys_world}. The phenomenon discovered long ago 
\cite{Hewson} has manifested itself  as a low temperature increase of the 
electrical resistance of diluted alloys. In quantum dots it  shows up as an 
increase of conductance  at low temperatures. The quantum dot devices allow the 
study of fundamental physics like Coulomb blockade phenomenon or Kondo effect 
in equilibrium and non-equilibrium conditions and in geometries not accessible 
in bulk systems. Both or one of the  external leads may be normal, 
superconducting or magnetic. In this paper we shall study the systems in which 
quantum dot is coupled with two ferromagnetic leads having the same or opposite 
magnetic polarization.

Spin polarized transport, especially single electron tunneling, in magnetic 
nanostructures has attracted much interest due to its potential applications 
in, for example, spintronics \cite{Prinz} and quantum computing \cite{Recher}. 
The Kondo effect in quantum dots attached to normal leads (N-QD-N structure) 
has been extensively studied both experimentally 
\cite{Goldhaber-Gordon}-\cite{Franceschi} and theoretically 
\cite{Glazman,Meir}. Many new effects have been predicted and observed in the 
transport characteristics, like splitting of the zero bias resonance under 
magnetic field \cite{Meir,Goldhaber-Gordon}, absence of the even-odd parity 
effects \cite{Schmid} or out of equilibrium Kondo effect 
\cite{Simmel,MK_1,Franceschi}. The Kondo effect has also been observed in many 
other systems: single atom \cite{Park}, single molecule \cite{Liang} and carbon 
nanotubes \cite{Nygard}. It has also been demonstrated in quantum dots attached 
to ferromagnetic leads \cite{Pasupathy}, where transport properties can, in 
principle, be  controlled with aid of the electron spin degree of freedom. 

Recently we observe  growing interest in electronic transport properties of the 
Kondo correlated quantum dots coupled to ferromagnetic electrodes 
\cite{Sergueev}-\cite{Utsumi}. In such geometry the Kondo resonance splits in 
parallel configuration \cite{Martinek_1, Dong_1, Martinek_2, Choi} (however, 
some of the works \cite{Zhang, Lopez} predict no splitting), while in the 
anti-parallel configuration the Kondo effect remains virtually the same as for 
nonmagnetic electrodes. The shot noise  studies \cite{Lu, Lopez} reveal huge 
differences for spin up and spin down electrons in the parallel alignment and 
no differences in anti-parallel configuration. 

It is well known that thermoelectric properties are the source of information 
complementary to that obtained from other transport characteristics 
\cite{Heremans}. Thermal properties (thermopower and thermal conductance) of 
the quantum dot coupled to the normal  leads in the Kondo regime have recently 
been investigated \cite{Boese}-\cite{Kim}. Thermopower has been shown to be 
very sensitive and powerful tool to study the Kondo effect. It manifests itself 
as an energy peak in the $DOS$ slightly below Fermi energy and this leads to 
change of sign of the thermopower.

It is the purpose of the present work to study the thermoelectric properties of 
the quantum dot coupled to ferromagnetic leads. We shall concentrate on the 
conductance, thermal conductance, thermopower and related quantities like 
tunnel magnetoresistance (TMR), thermoelectric figure of merit which value is 
direct sign of usefulness of the system for applications and Widemann-Franz 
ratio, which normalized value differing from 1 signals breakdown of the Fermi 
liquid state.

We show that thermopower is very different for spin up and spin down electrons 
in parallel configuration and is similar to the nonmagnetic case for 
anti-parallel one. However, for parallel alignment the total (spin up plus spin 
down) thermopower is very small compared to the anti-parallel alignment. 

The organisation of the rest of the paper is as follows. In Ssetion II we
present the model and discuss some aspects of calculational procedure.
Results of calculations are presented and discussed in Section III. We end up
with summary and conclusions.


\section{Model and approach}
\begin{figure}[h]
 \resizebox{0.7\linewidth}{!}{ \includegraphics{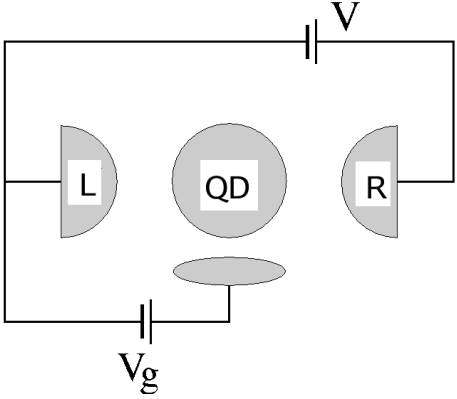}}
  \caption{\label{Fig1} Schematic view of the quantum dot coupled to two leads 
           R and L, which may be magnetically polarized and/or at different 
	   temperatures and voltages. Voltage applied to the junction is taken 
	   into consideration as a shift of the chemical potentials 
	   $\mu_{L}-\mu_{R}=eV$.}
\end{figure}

Schematic view of the quantum dot coupled to two leads R and L, which may be 
magnetically polarized and/or at different temperatures and voltages is shown 
in the Fig. \ref{Fig1}.

We assume that the interaction energy between two electrons on the quantum dot 
is the largest energy in the problem and thus model the system as the 
$U = \infty$ single impurity Anderson Hamiltonian \cite{Anderson} in the slave 
boson representation \cite{LeGuillou}-\cite{MK_3}
\begin{eqnarray}
H = \sum_{\lambda {\bf k} \sigma} \epsilon_{\lambda {\bf k} \sigma} 
c^+_{\lambda {\bf k} \sigma} c_{\lambda {\bf k} \sigma} 
+ \sum_{\sigma} \varepsilon_{\sigma} f^+_{\sigma} f_{\sigma} \nonumber \\
+ \sum_{\lambda {\bf k} \sigma} \left( V_{\lambda {\bf k} \sigma} 
c^+_{\lambda {\bf k} \sigma} b^+ f_{\sigma} + H. c. \right)
\label{Hamilt}
\end{eqnarray}
where $\lambda = L$ ($R$) denotes left (right) lead, 
$c^+_{\lambda {\bf k} \sigma}$ ($c_{\lambda {\bf k} \sigma}$) is the creation 
(annihilation) operator for a conduction electron with the wave vector 
${\bf k}$, spin $\sigma$ in the lead $\lambda$ and $V_{\lambda {\bf k} \sigma}$ 
is the hybridization matrix element between localized electron on the dot with 
the energy $\varepsilon_{\sigma}$ and conduction electron of energy 
$\epsilon_{\lambda {\bf k}}$ in the lead $\lambda$. Ferromagnetism of the 
electrodes is modeled via modified conduction electron energy 
$\epsilon_{\lambda {\bf k} \sigma} = \epsilon_{\lambda {\bf k}} \pm \sigma h_z$ 
where the magnetization points into $z$ direction (the same or opposite in both 
leads).

The particle current $J_{\lambda}$ and the energy flux $J_{E\lambda}$ flowing 
from the lead $\lambda$ to the central region can be calculated from the time 
derivative of charge and energy operator respectively \cite{Haug}. We use 
relation $J_{Q\lambda} = J_{E\lambda} - \mu_{\lambda} J_{e \lambda}$ for the 
thermal flux $J_{Q\lambda}$ and express all currents in terms of Keldysh Green 
functions \cite{Keldysh} in the standard form \cite{Boese,Dong_1}
\begin{eqnarray}
J_{e \lambda} = \frac{i e}{\hbar} \sum_{\sigma} \int^{\infty}_{-\infty}
\frac{d\omega}{2 \pi} \Gamma_{\lambda \sigma}(\omega) 
[ G^<_{\sigma}(\omega) + 
2 i f_{\lambda}(\omega) {\rm Im} G^r_{\sigma}(\omega) ]
\label{p_curr}
\end{eqnarray}
\begin{eqnarray}
J_{Q \lambda} = \frac{i}{\hbar} \sum_{\sigma} \int^{\infty}_{-\infty}
\frac{d\omega}{2 \pi} \Gamma_{\lambda \sigma}(\omega) (\omega - 
\mu_{\lambda}) 
[ G^<_{\sigma}(\omega) \nonumber \\
+ 2 i f_{\lambda}(\omega) {\rm Im} G^r_{\sigma}(\omega) ]
\label{h_curr}
\end{eqnarray}
where $G^r_{\sigma}(\omega)$ is the Fourier transform of the retarded Green 
function (GF) 
$G^r_{\sigma}(t, t') = 
i \theta \langle [f_{\sigma}(t), f^+_{\sigma}(t')]_+ \rangle$ and 
$G^<_{\sigma}(\omega) = i \langle f^+_{\sigma}(t') f_{\sigma}(t) \rangle$ is 
the Fourier transform of the lesser Keldysh GF \cite{Keldysh}. 
$\Gamma_{\lambda \sigma}(\omega) = 
2 \pi \sum_{\bf k} |V_{\lambda {\bf k} \sigma}|^2 
\delta(\omega - \epsilon_{\lambda {\bf k} \sigma})$ denotes the strength of the 
coupling between dot and the lead $\lambda$, 
$f_{\lambda}(\omega) = f(\omega - \mu_{\lambda})$ is the Fermi distribution 
function in the lead $\lambda$ with the chemical potential $\mu_{\lambda}$ and 
temperature $T_{\lambda}$. 

In general, when both the strong on-dot Coulomb interaction and the tunneling 
between dot and leads occur, it is not possible to calculate 
$G^{<(r)}_{\sigma}(\omega)$ exactly. Several approximation schemes have been 
proposed to calculate $G^{<(r)}_{\sigma}(\omega)$. Here we use recently 
proposed equation of motion technique for nonequilibrium GF \cite{Niu}. This 
technique allows  calculation of both  $G^<_{\sigma}(\omega)$ and 
$G^r_{\sigma}(\omega)$ in consistent way making similar approximations in the 
decoupling procedure of both $G^<_{\sigma}(\omega)$ and $G^r_{\sigma}(\omega)$. 
The approach has been successfully applied in quantum dot systems 
\cite{MK_2, MK_3}. In the present case it yields 
\begin{eqnarray}
J_e = -\frac{e}{h} \sum_{\sigma} \int^{\infty}_{-\infty}
d\omega \Gamma_{\sigma}(\omega) 
[f_L(\omega) - f_R(\omega)] {\rm Im} G^r_{\sigma}(\omega) 
\label{pf_curr}
\end{eqnarray}
\begin{eqnarray}
J_Q = -\frac{1}{h} \sum_{\sigma} \int^{\infty}_{-\infty}
d\omega \Gamma_{\sigma}(\omega) (\omega - eV) \nonumber \\ 
\times 
[f_L(\omega) - f_R(\omega)] {\rm Im} G^r_{\sigma}(\omega)
\label{hf_curr}
\end{eqnarray}
where $\Gamma_{\sigma} = \Gamma_{L \sigma} \Gamma_{R \sigma}/ 
[\Gamma_{L \sigma} + \Gamma_{R \sigma}]$ and $eV = \mu_L - \mu_R$. 
The on-dot retarded GF reads
\begin{eqnarray}
G_{\sigma}(\omega) = 
\frac{1 - \langle n_{-\sigma} \rangle}{\omega - \varepsilon_{\sigma} - 
\Sigma_{0 \sigma}(\omega) - \Sigma_{I \sigma}(\omega)}
\label{GF}
\end{eqnarray}
with noninteracting 
$\Sigma_{0 \sigma}(\omega) = \sum_{\lambda {\bf k}} 
\frac{|V_{\lambda {\bf k} \sigma}|^2}{\omega-
\epsilon_{\lambda {\bf k} \sigma}}$
and interacting self-energy 
$\Sigma_{I \sigma}(\omega) = \sum_{\lambda {\bf k}} 
\frac{|V_{\lambda {\bf k} -\sigma}|^2 
f_{\lambda}(\epsilon_{\lambda {\bf k} -\sigma})}
{\omega - (\varepsilon_{\sigma}-\varepsilon_{-\sigma}) - 
\epsilon_{\lambda {\bf k} -\sigma}}$. 
In order to get the splitting of the Kondo resonance in the presence of the 
ferromagnetic leads, which is consistent with the scaling analysis, we follow 
Ref. \cite{Martinek_1} and replace the bare dot energy level 
$\varepsilon_{\sigma}$ by $\tilde \varepsilon_{\sigma}$ found self-consistently 
from $\tilde \varepsilon_{\sigma} = \varepsilon_{\sigma} + 
{\rm Re} [\Sigma_{0\sigma}(\tilde \varepsilon_{\sigma}) + 
\Sigma_{I\sigma}(\tilde \varepsilon_{\sigma})]$ 
in the interacting self-energy 
$\Sigma_{I\sigma}(\omega)$. 
In numerical results presented below we have used constant bands of width 
$D = 100 \Gamma$, $\frac{1}{2}(\Gamma_{L\sigma}+\Gamma_{R\sigma}) = \Gamma$ and 
use $\Gamma$ as our energy unit in the following.

Within linear response theory for the particle current and the heat flux one  
defines the conductance $G$ as equal to $-\frac{e^2}{T} L_{11}$, thermopower is 
given by $S = -\frac{1}{eT} \frac{L_{12}}{L_{11}}$ and the thermal conductance 
by $\kappa = \frac{1}{T^2} \left(L_{22} - \frac{L^2_{12}}{L_{11}}\right)$. The 
linear response coefficients read 
\begin{eqnarray}
L_{11} = \frac{T}{h} \sum_{\sigma} \int d\omega \Gamma_{\sigma} (\omega) 
{\rm Im} G^r_{\sigma}(\omega) 
\left(-\frac{\partial f(\omega)}{\partial \omega} \right)_T
\label{L11}
\end{eqnarray}
\begin{eqnarray}
L_{12} = \frac{T^2}{h} \sum_{\sigma} \int d\omega \Gamma_{\sigma} (\omega) 
{\rm Im} G^r_{\sigma}(\omega) 
\left(\frac{\partial f(\omega)}{\partial T} \right)_{\mu}
\label{L12}
\end{eqnarray}
\begin{eqnarray}
L_{22} = \frac{T^2}{h} \sum_{\sigma} \int d\omega \Gamma_{\sigma} (\omega) 
(\omega - eV) {\rm Im} G^r_{\sigma}(\omega) 
\left(\frac{\partial f(\omega)}{\partial T} \right)_{\mu}
\label{L22}
\end{eqnarray}
%


The leads considered here are magnetically polarized. In the following we shall 
characterize the degree of polarization by  the parameter 
$p = \frac{\langle n_{\uparrow} \rangle - \langle n_{\downarrow} \rangle}
{\langle n_{\uparrow} \rangle + \langle n_{\downarrow} \rangle}$, where 
$n_{\sigma}$ is the concentration of spin $\sigma$ electrons. The magnetization 
of both leads may point into the same direction (parallel configuration) or in 
opposite directions (anti-parallel configuration). 

\section{The results}
\begin{figure}[h]
 \resizebox{0.8\linewidth}{!}{
  \includegraphics{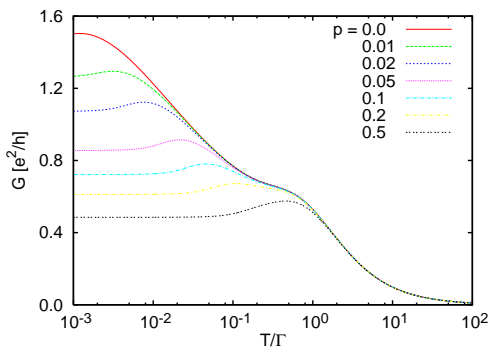}}
  \caption{\label{Fig2} Temperature dependence of the conductance for a number 
           of leads polarizations. $\varepsilon_d = -1.75 \Gamma$. }
\end{figure}

In  Fig.\ref{Fig2} linear conductance of the system in parallel configuration 
is calculated as a function of temperature for a number of leads polarizations
$p$.
One observes the decrease of $G$ for larger values of $p$. This is due to the
splitting of the Kondo resonance. Moreover, the decrease of $G$ is not 
universal function of $p$. On the other hand, in anti-parallel configuration,
the conductance can be obtained from that one for unpolarized leads, as it 
scales according to the relation: $G(p)/G(0) = 1 - p^2$ for the whole
temperature region. 

\begin{figure}[h]
 \resizebox{0.8\linewidth}{!}{
  \includegraphics{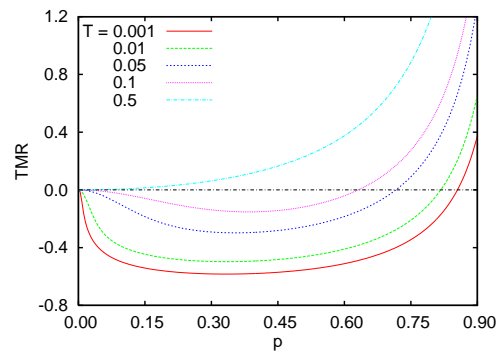}}
  \caption{\label{Fig3} TMR as a function of the polarization factor $p$ for a 
           number of temperatures. $\varepsilon_d=-1.75 \Gamma$, 
	   $T_K(p=0) \approx 2\cdot 10^{-2}\Gamma$.}
\end{figure}

Tunnel magnetoresistance is defined \cite{Lopez2} as the ratio  
\begin{eqnarray}
TMR=\frac {G_P-G_{AP}}  {G_{AP}}
\nonumber
\end{eqnarray}
where $G_{P(AP)}$ is the conductance calculated for parallel (anti-parallel) 
configuration of the leads magnetization. It is shown in Fig. \ref{Fig3} as a 
function of polarization factor p for a number of temperatures for system 
characterized by $\varepsilon_d=-1.75 \Gamma$. 
Note that the Kondo temperature itself is a function of $p$. For non-polarized 
leads one finds $T_K(p=0) \approx 2\cdot 10^{-2}\Gamma$. The dependence is not 
symmetric with respect to $p=0.5$ and changes its character with temperature.
For small polarizations it is negative, as in this case $G_{AP}$ is not much
affected by the exchange field while $G_P$ is strongly suppressed due to the
splitting of the Kondo resonance. For large values of the polarization, TMR is
positive, as in this case $G_{AP}$ goes to zero with $p \rightarrow 1$ while
$G_P$ tends to finite value. 

\begin{figure}[h]
 \resizebox{0.8\linewidth}{!}{
  \includegraphics{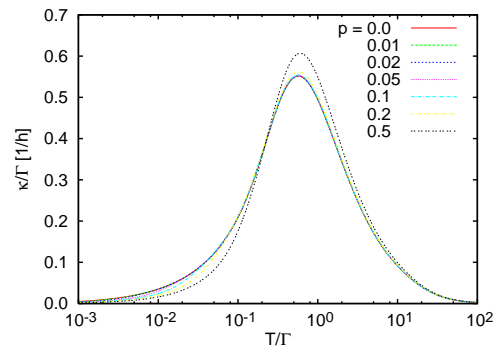}}
 \caption{\label{Fig4} Thermal conductance of the 
 quantum dot as a function of
 temperature for a number of lead polarizations.}
\end{figure}

The thermal conductance $\kappa$ vs. temperature in the $P$ configuration is 
displayed in the Fig. \ref{Fig4}.
As one can see in the figure \ref{Fig4} thermal conductance does not change 
much with the increasing of the leads polarization $P$. This quantity is not 
sensitive tool for the study of the Kondo effect. In the case of $AP$
configuration for any temperature $\kappa$ scales with the polarization in the 
same way as $G$ does, namely, $\kappa(p)/\kappa(0) = 1 - p^2$.

\begin{figure}[h]
 \resizebox{0.8\linewidth}{!}{
  \includegraphics{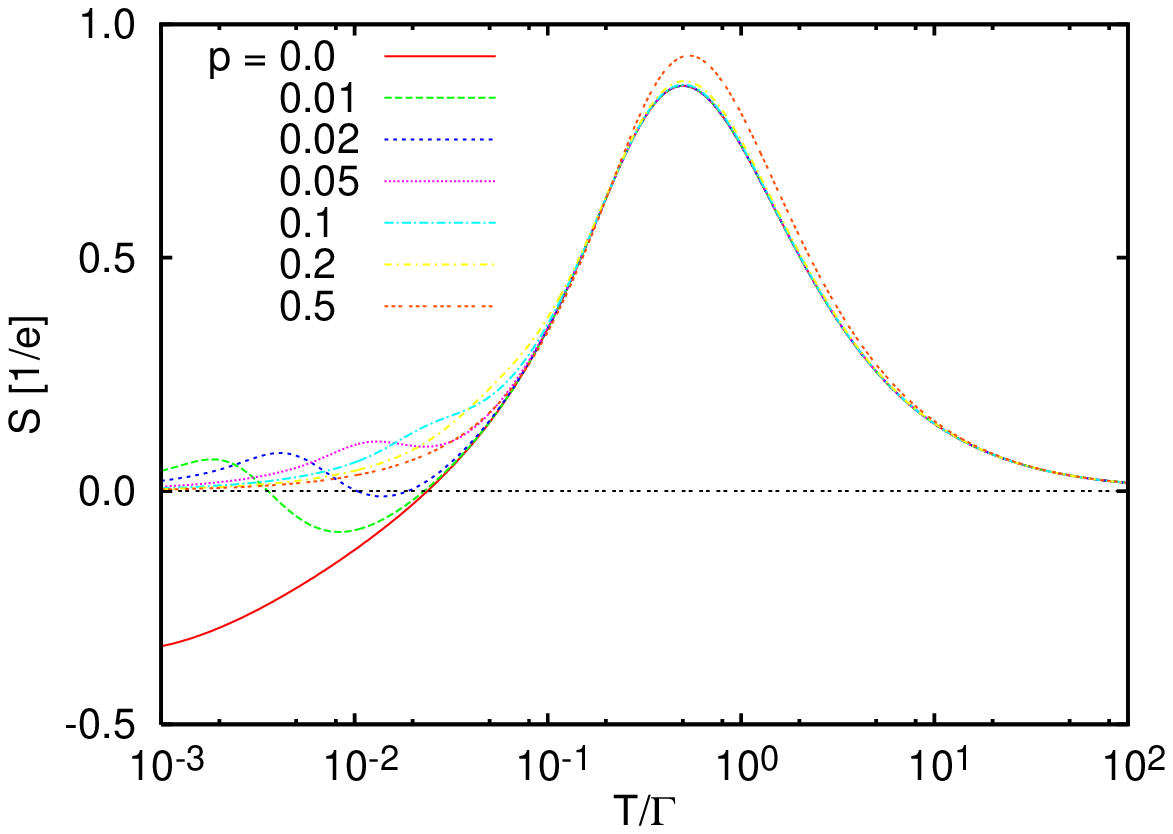}}
 \resizebox{0.8\linewidth}{!}{
  \includegraphics{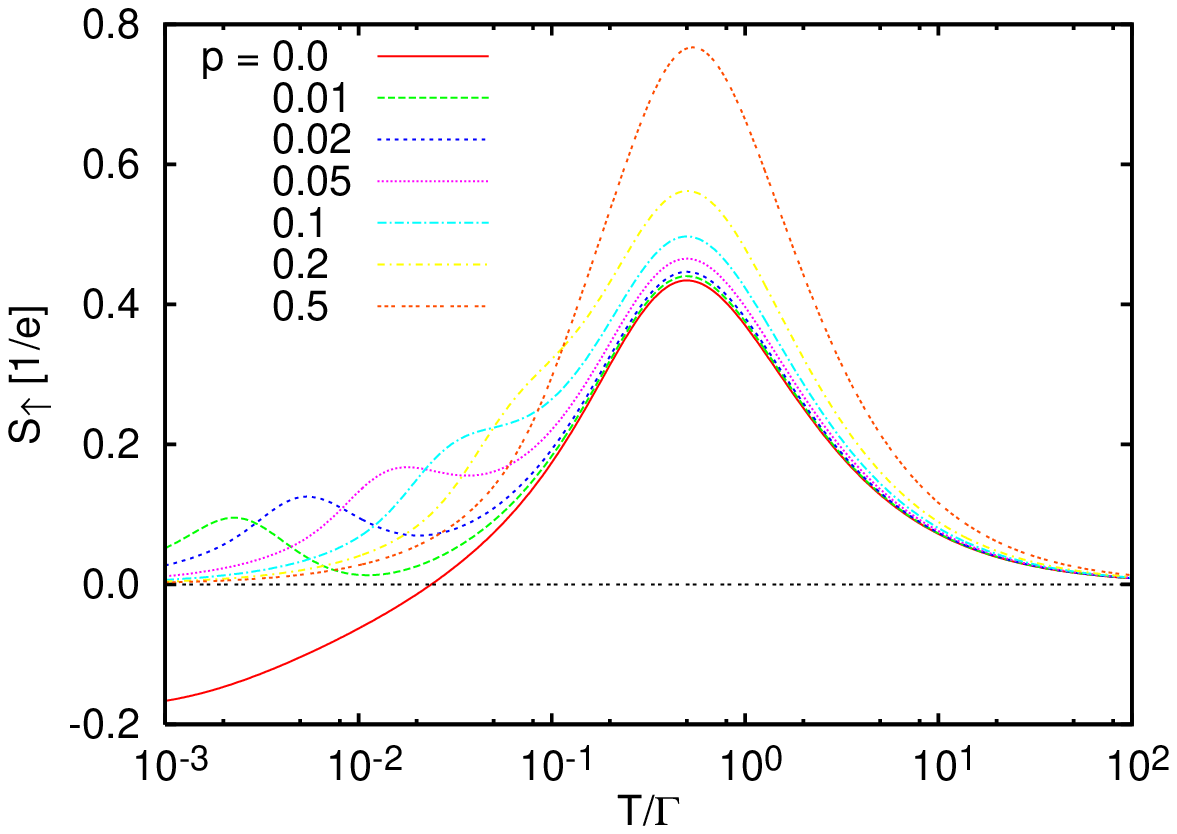}}
 \resizebox{0.8\linewidth}{!}{
  \includegraphics{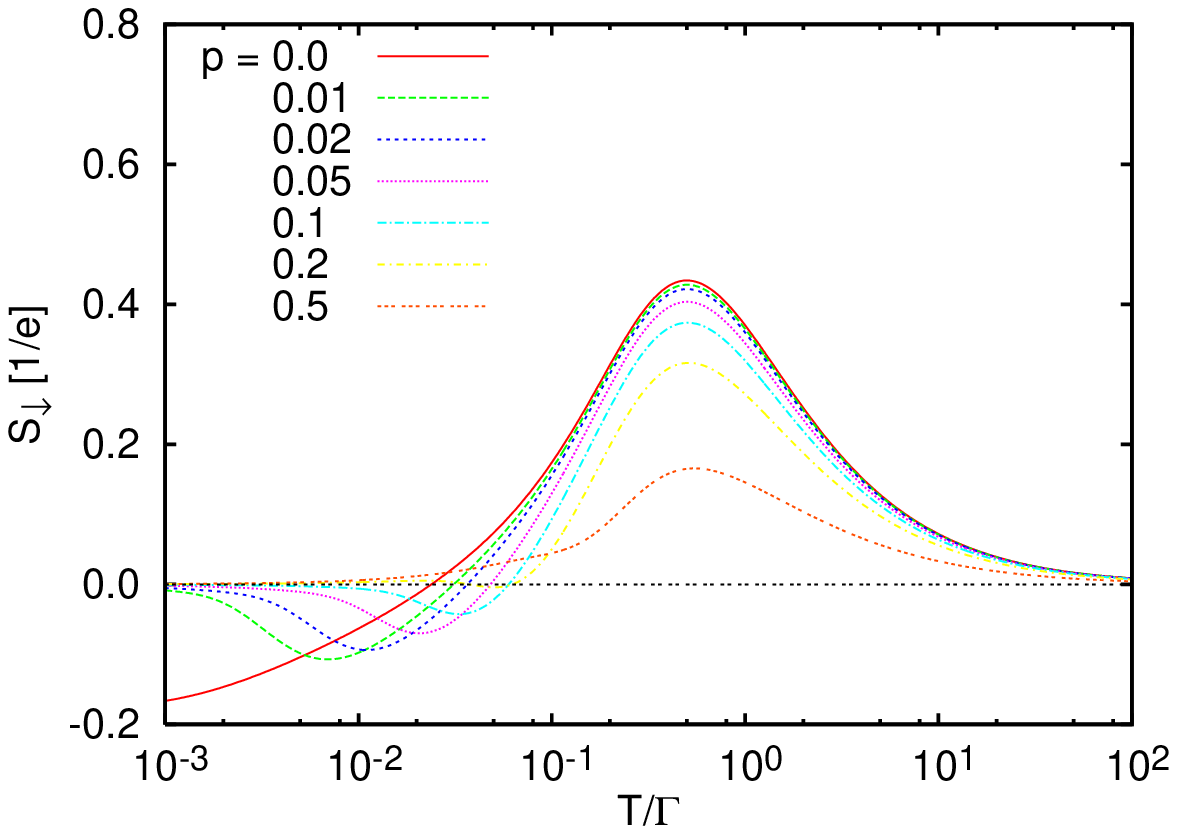}}
 \caption{\label{Fig5} The total thermopower of the quantum dot (upper panel), 
          the thermopower associated with the spin up electrons (middle panel) 
	  and that one associated with spin down electrons (lower panel) as a 
	  function of the temperature for a number of lead polarizations.}
\end{figure}

In  the Fig. \ref{Fig5} linear thermopower $vs.$ temperature is shown for a 
number of polarizations in the parallel configuration. 

The upper panel shows the total thermopower, the middle one the thermopower 
associated with spin up while lower one with spin down electrons. The $p=0$ 
curve corresponds to non-magnetic leads. First of all one can see that for 
$p=0$ and around $T=T_K$, where $T_K$ is the Kondo scale, the thermopower 
reaches minimal value. It increases for elevated temperatures and eventually 
changes sign. This signals the disappearance of the Kondo peak. As it is well 
known, and in similar context has already been noted by Boese and Fazio 
\cite{Boese}, the thermopower is sensitive to the curvature of the energy 
dependence of the density of states. This curvature is negative at high 
temperatures and for Fermi level above the on-dot energy level, while it 
becomes negative at low temperatures, when the Kondo resonance forms slightly 
above Fermi level.

The low temperature changes of $S$ with polarization can also be understood. 
With increasing polarization in the leads thermopower decreases to zero, except 
at very high temperatures where the broad maximum (minimum) is observed. Small 
thermopower at low $T$ for increasing leads polarization is due to the 
splitting of the Kondo resonance by the stray fields coming from the 
ferromagnetic leads. In this case DOS around the Fermi energy is almost 
symmetric. Moreover the thermopower is mostly positive in the whole range of 
the temperatures except for very small leads polarizations. It is worthwhile to 
note additional broad maximum (spin up) and minimum (spin down) of the 
thermopower at temperatures below  $T = 0.1$ for small polarizations. For such 
polarizations the distance between Kondo resonances $\delta$ is not so large 
($\delta \approx 1$) and thermal broadening significantly affects the DOS at 
the Fermi level, thus giving rise to the observed thermopower.
 
For $AP$ configuration the thermopower does not depend on the polarization and
is the same as for the QD with non-magnetic leads. This can be easily 
understood as this quantity measures the curvature of the DOS around the Fermi
energy, which does not change with the polarization in this case. Only
the height of the Kondo resonance changes.   

\begin{figure}[h]
 \resizebox{0.8\linewidth}{!}{
  \includegraphics{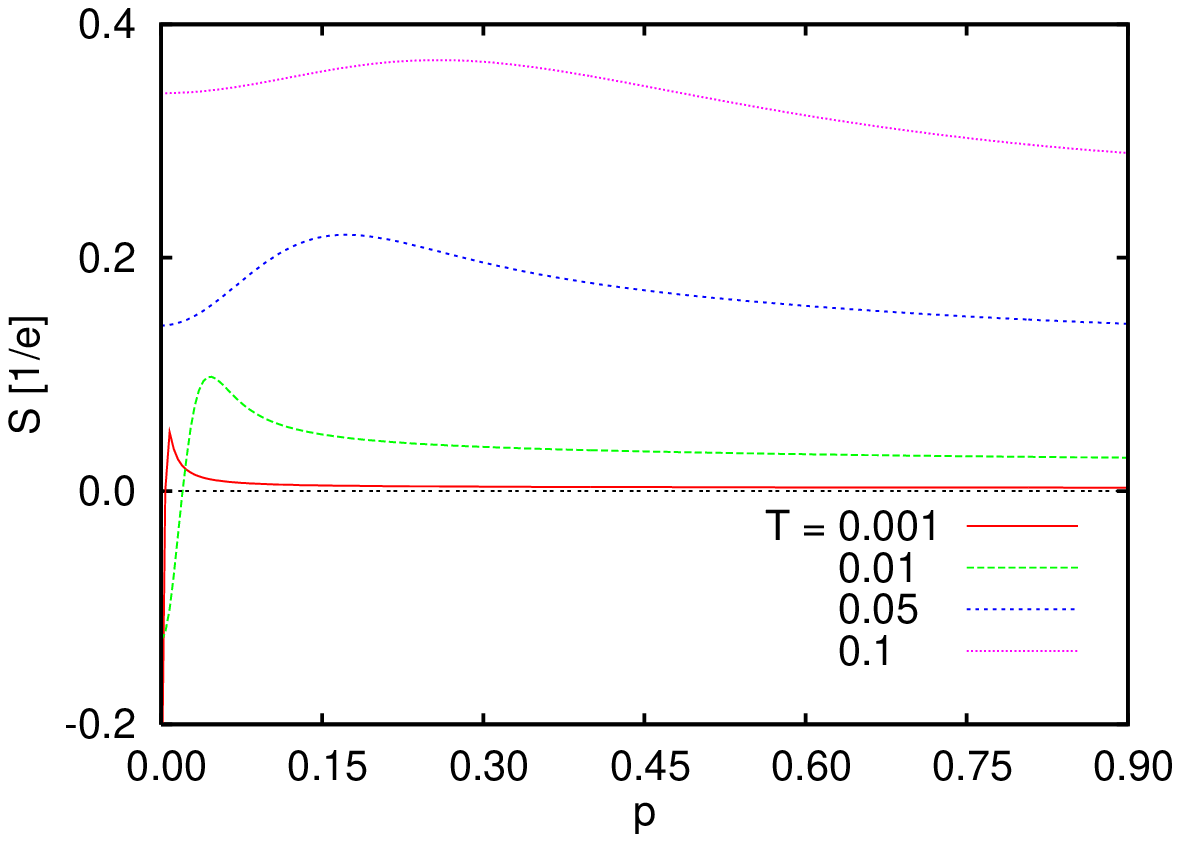}}
 \resizebox{0.8\linewidth}{!}{
  \includegraphics{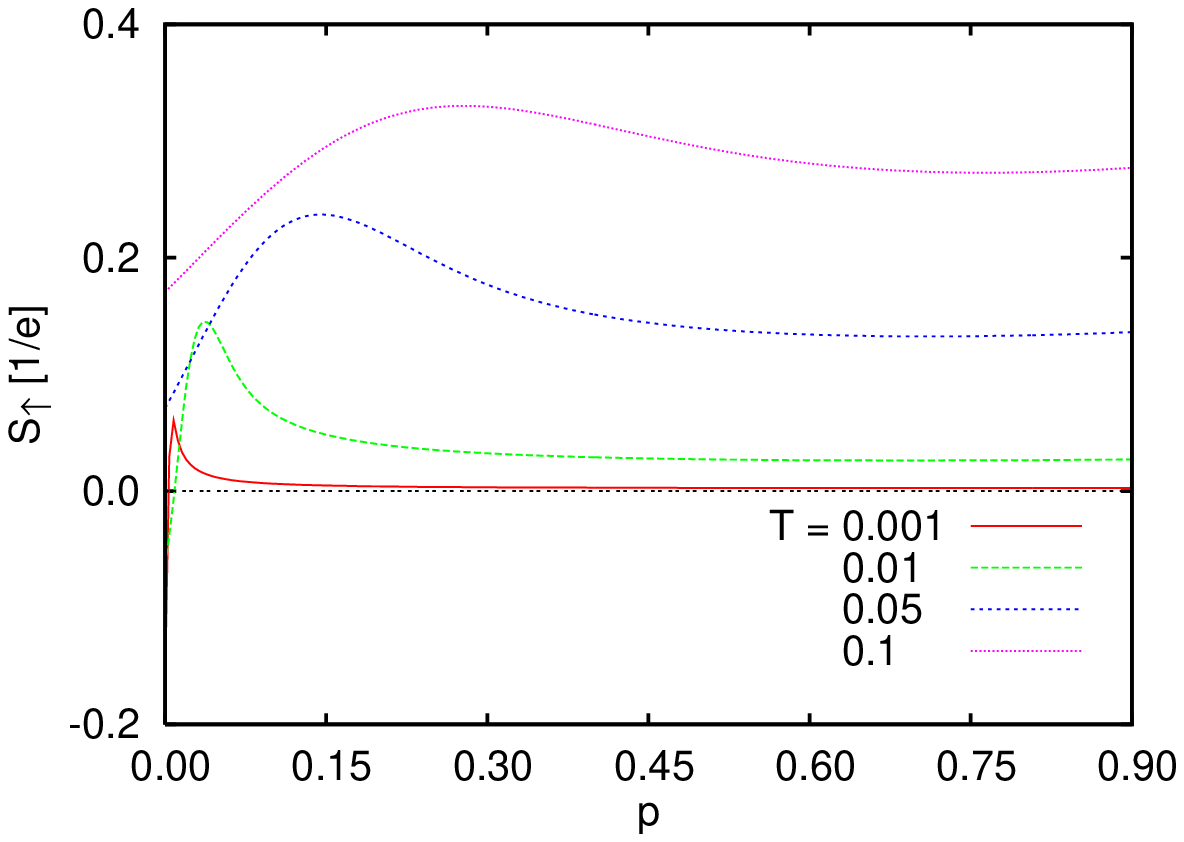}}
 \resizebox{0.8\linewidth}{!}{
  \includegraphics{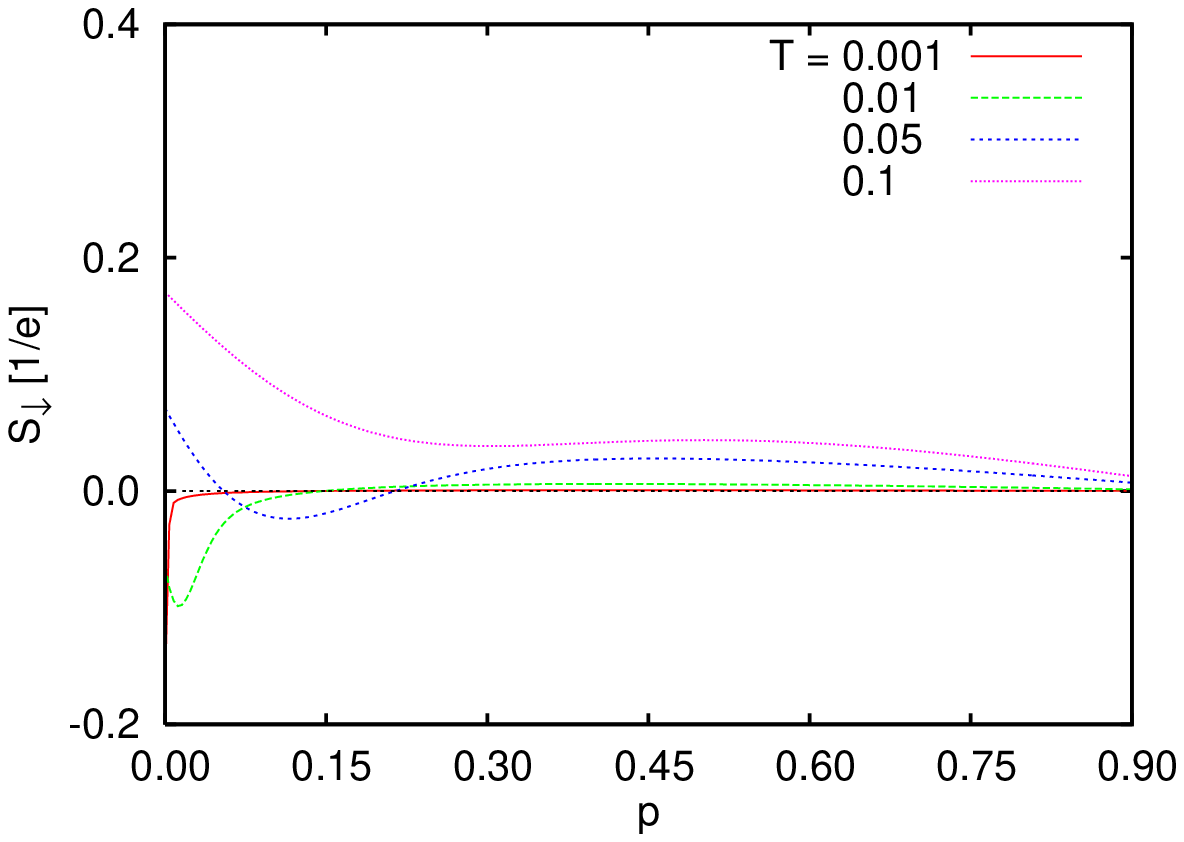}}
 \caption{\label{Fig6} Thermopower of the quantum dot as a function of the lead 
          polarization for a number of temperatures. The upper panel shows the
	  total thermopower, the middle one - that associated with spin up
	  electrons, and the lower one - that for spin down electrons. }
\end{figure}

In the Fig. \ref{Fig6} the total thermopower (upper panel), 
that  in the spin
up channel (middle panel), and in the spin down channel (lower panel), is 
plotted as a function of the lead polarization in $P$ configuration.
Interestingly thermopower changes its behavior in significant way only for 
small values of $p$ and at low temperatures it even changes sign. This is 
attributed to the splitting of the Kondo resonances. For small lead 
polarizations the splitting is small but significantly influences the DOS 
around the Fermi energy. For larger $p$ the Kondo peaks are far away from the 
Fermi energy and do not change low energy DOS much.

So far the results for parallel configuration of the leads polarizations have 
been presented. However, as we have mentioned already, for anti-parallel 
configuration the calculated quantities either do not depend on the value of 
the polarization and they are the same as for non-magnetic leads (thermopower)
or can be obtained from the solution for non-magnetic leads due to the 
scaling relation $1 - p^2$ (electric and thermal conductance). This is easy to 
understand as for such polarizations the tunneling of spin up and spin down 
electrons on the dot are allowed to one of the electrodes. This is true for 
symmetric couplings. The situation is different when there is asymmetry in the 
coupling to the left and right lead. In this case the electric conductance
behaves similarly as in parallel configuration (see Fig. \ref{Fig2}). Thermal 
conductance also does not change much with increasing of the lead polarization 
but unlike for $P$ configuration, where it decreases (increases) at low (high) 
temperature, it decreases in the whole region of the temperatures (see
discussion below the Fig. \ref{Fig4}). 

\begin{figure}[h]
 \resizebox{0.8\linewidth}{!}{
  \includegraphics{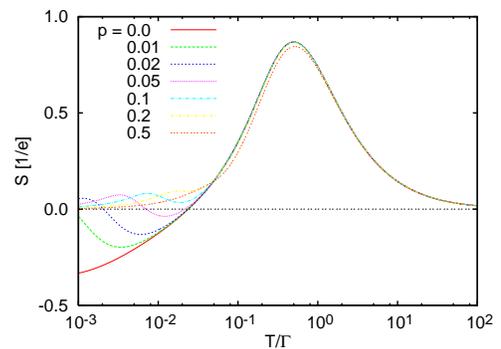}}
 \caption{\label{Fig7} Thermopower vs. temperature for asymmetrically 
          ($\Gamma_{L\sigma}/\Gamma_{R\sigma} = 2$) coupled quantum dot in the
	  anti-parallel configuration.}
\end{figure}

On the other hand, behavior of the thermopower in asymmetrical $AP$ 
configuration (which does not depend on $p$ for symmetric couplings) is similar 
to the case of the $P$ configuration (see Fig. \ref{Fig5}). In the Fig. 
\ref{Fig7} thermopower vs. temperature for asymmetrically coupled quantum dot 
with $\Gamma_{L\sigma}/\Gamma_{R\sigma} = 2$ is plotted.
At high T thermopower goes to zero, eventually oscillating, with decreasing 
temperature. 

Such behavior can be explained by the fact that for asymmetric couplings there 
is a splitting of the Kondo resonance. In symmetrically coupled quantum dot 
there is an equal number of electrons with spin up and down on the dot coming 
from different leads and one can show that this model can be mapped onto 
quantum dot with non-magnetic leads. Here, when 
$\Gamma_{L\sigma}/\Gamma_{R\sigma} \neq 1$, due to different tunneling 
probabilities, quantum dot sees asymmetry in electron number with spins up and 
down and in this sense anti-parallel configuration is very similar to parallel
one (see Fig. \ref{Fig5}). 

\begin{figure}[h]
 \resizebox{0.8\linewidth}{!}{
  \includegraphics{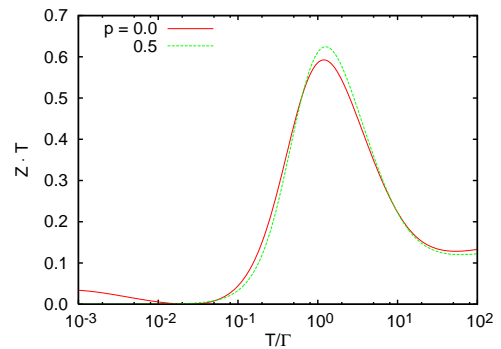}} 
 \caption{\label{Fig8} Temperature dependence of the thermoelectric figure of 
          merit $Z \cdot T =S^2 G T / \kappa$.}
\end{figure}

Thermoelectric figure of merit $Z  = S^2 G / \kappa$ is a direct measure of the 
usefulness of the material or device for thermoelectric power generators or 
cooling systems \cite{Heremans}. For simple systems it is inversely 
proportional to operation temperature and thus one conveniently plots 
$Z \cdot T$, which numerical value is an indicator of the systems performance. 
In the Fig. \ref{Fig8} we show $Z \cdot T$ as a function of temperature in the
$P$ configuration and note that it is smaller than 1, which signals limited 
applicability of the studied device. 
Note that polarization slightly increases $Z \cdot T$ but it never exceeds 1. 

\begin{figure}[h]
 \resizebox{0.8\linewidth}{!}{
  \includegraphics{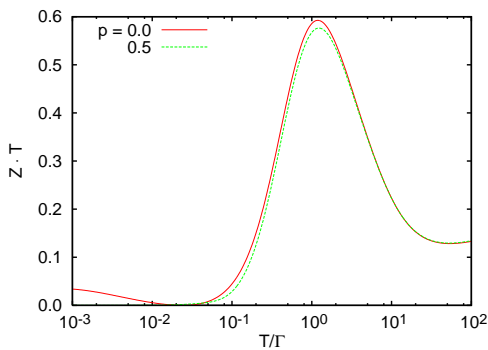}}
  \caption{\label{Fig9} Temperature dependence of the thermoelectric figure of 
           merit $Z \cdot T =S^2 G T / \kappa$ of the asymmetrically coupled
	   quantum dot with $\Gamma_{L\sigma}/\Gamma_{R\sigma} = 2$ in 
	   anti-parallel configuration.}
\end{figure}

Again, in $AP$ configuration the thermoelectric figure of merit does not depend
on $p$, which can be easily deduced from the definition of this quantity. On 
the other hand, in asymmetrical $AP$ configuration it changes little but unlike
for $P$ configuration, it decreases with increasing of the polarization. This is
shown in the Fig. \ref{Fig9}. 

\begin{figure}[h]
 \resizebox{0.8\linewidth}{!}{
  \includegraphics{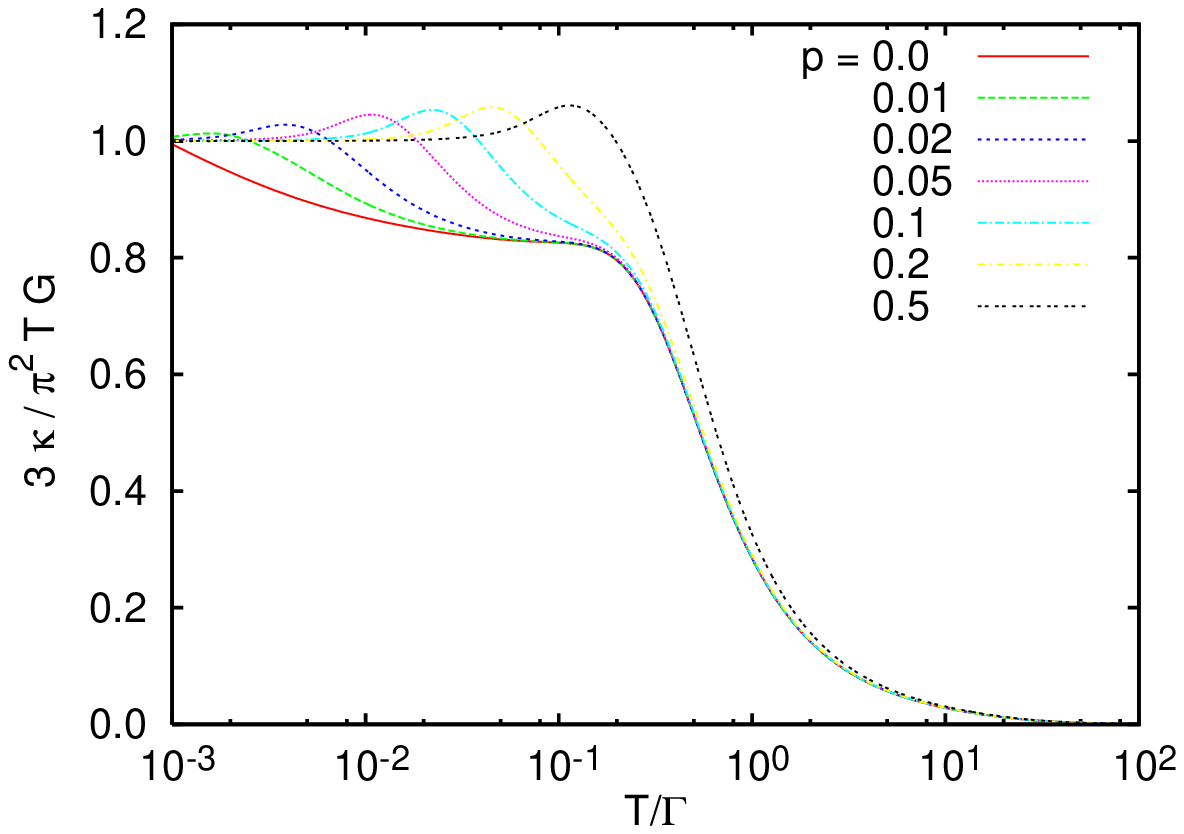}}
 \resizebox{0.8\linewidth}{!}{
  \includegraphics{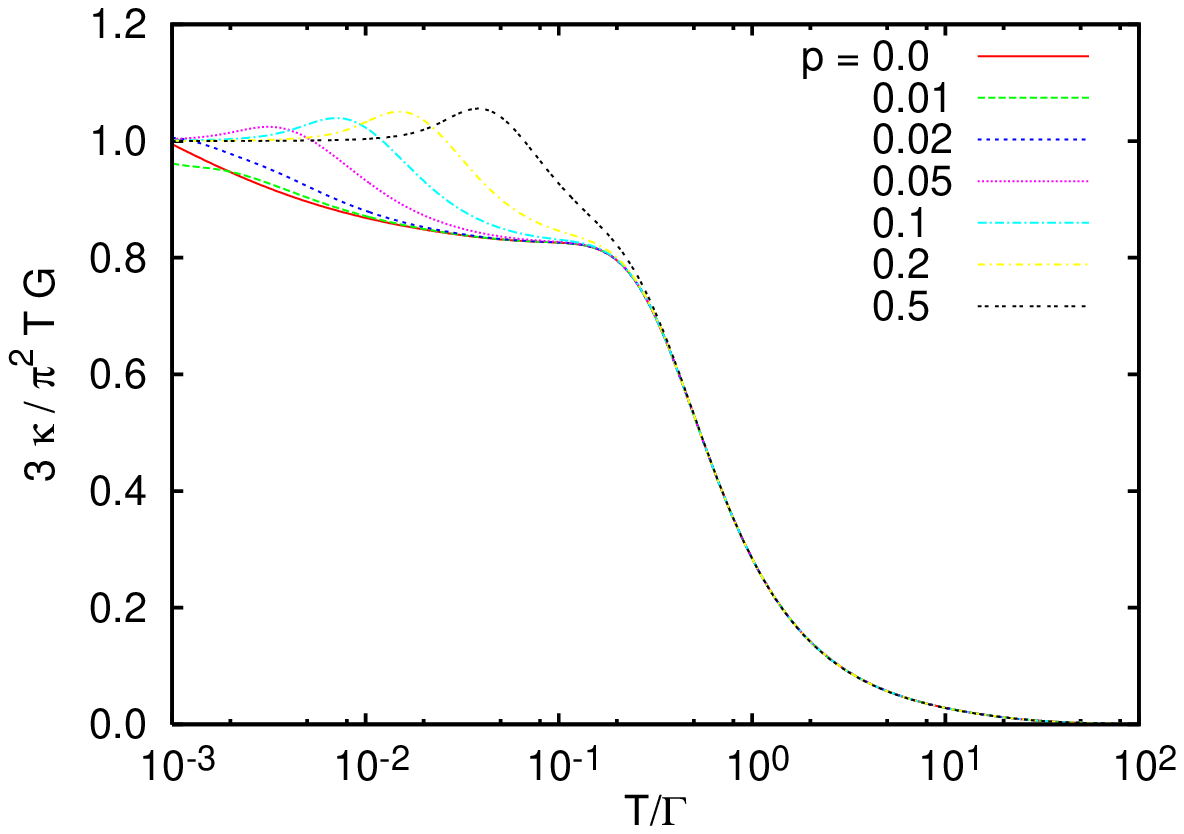}}
  \caption{\label{Fig10}  Temperature dependence of the  Widemann-Franz ratio 
           $L/L_0= \frac{\pi^2}{3 e^2} (\kappa /T G)$, with 
	   $L_0= \frac{3 e^2}{\pi^2}$ as a function of temperature for a number
	   of the leads polarizations for symmetric coupling in $P$
	   configuration (upper panel) and for asymmetrically coupled dot with 
	   $\Gamma_{L\sigma}/\Gamma_{R\sigma} = 2$ in anti-parallel 
	   configuration (lower panel).}
\end{figure}

Finally we discuss Wiedemann-Franz (WF) law which relates thermal and 
electrical transport via relation $\kappa = \frac{\pi^2}{3 e^2} T G$. This law 
describes transport in Fermi liquid bulk metals and in general is not obeyed in 
QD systems where the transport occurs through small confined region 
\cite{Boese,Dong_2}. However at very low temperatures, where the Kondo effect 
develops and the ground state of the system has Fermi liquid nature, the WF law 
is recovered. At high temperatures transport is dominated by sequential 
tunneling processes leading to the larger suppression of the thermal transport 
than the electrical one. This behavior is illustrated in the Fig. (\ref{Fig10})
in $P$ configuration (upper panel) and in asymmetrical $AP$ configuration 
(lower panel) for a number of the polarizations.

In the $AP$ configuration with symmetric couplings to the leads WF ratio does
not depend on $p$ due to its definition and behavior of the quantities 
entering into it. 

\section{Comparison with experiment}

\begin{figure}[h]
\resizebox{0.8\linewidth}{!}
{\includegraphics[width=0.38\textwidth]{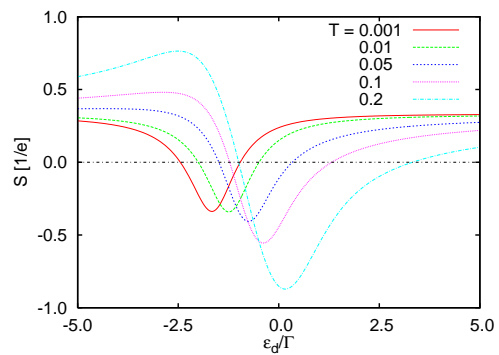}} 
\caption{\label{fig11}The linear response thermopower of the quantum dot 
         coupled to non-magnetic leads as a function of the position of the 
	 on-dot electron energy calculated for different temperatures (in units 
	 of $\Gamma$): $T=10^{-3}$, $10^{-2}$, $5 \cdot 10^{-2}$, $10^{-1}$ and
	 $2 \cdot 10^{-1}$.}  
\end{figure}

It is worth to note that the thermopower of the quantum dot in the Kondo regime 
has recently been measured experimentally \cite{Scheibner} olbeit for 
nonmagnetic electrodes. The dot containing few electrons was defined in the two 
dimensional electron gas in which also  the external electrodes were defined. 
In the experimental situation the charging energies ({\it i.e.} U) were finite 
and varied between 2 and 5 in units of effective coupling $\Gamma$. This 
precludes the direct comparison with our calculations (for p=0)  as we have 
taken $U=\infty$ limit. Experimentally one applies the bias $V_{SD}$ between 
source and drain and measures thermopower as function of gate voltage $V_g$. 
Theoretically this dependence can be modeled by plotting $S$ as function of 
on-dot energy level $\varepsilon_d$. Experimental data, shown in figure 4 of 
reference \cite{Scheibner} qualitatively agree with our calculations for $p=0$ 
presented in figure \ref{fig11}. 

To understand the  results, as the first approximation one can use so called 
Mott formula, which states that thermopower is proportional to the logarithmic 
derivative of the conductance with respect to the energy evaluated at the 
actual Fermi energy. The Kondo effect shows up as additional maximum appearing 
in conductance $via$ quantum dot when the temperature decreases. It is 
essentially due to Abrikosov-Suhl resonance in the density of states, which 
appears at low temperatures and is located slightly above the Fermi level of 
the leads. The slope of conductance thus changes and the thermopower changes 
sign. The similarity between theoretical Fig.\ref{fig11} and experimental 
Fig. 4 of \cite{Scheibner} results is very encouraging. However for qualitative 
comparison one has to take few energy levels on the dot, assume finite $U$ 
values and selconsistently calculate the shift of $\varepsilon_d$ with changing 
gate voltage and  nonlinear conductance and thermopower for actual value of 
$V_g$ and $V_{SD}$. This is outside the scope of the present paper. The 
detailed comparison between experimental data and calculations will be the 
subject of future work. 

\section{Summary and Conclusions}
In summary, we have studied thermal properties of the strongly correlated 
quantum dot coupled to the ferromagnetic leads in the Kondo regime. We have 
found that thermopower is strongly suppressed at low temperatures due to the 
splitting of the Kondo resonance in parallel configuration of the lead 
polarization. In anti-parallel configuration Kondo effect behaves in the way 
similar to the system with non-magnetic electrodes so the results do not depend 
on the value of the polarization in the leads. Moreover we have shown that 
asymmetry in the coupling to the leads in anti-parallel configuration has 
important consequences as it lifts spin degeneracy on the dot thus leading to 
the suppression of the thermopower at low temperature, similarly as in the 
parallel polarization configuration. Finally we have checked the 
Wiedemann-Franz relation which does not hold in general, but similarly as for 
QD with non-magnetic leads it is recovered at low temperatures where the Kondo 
effect develops. The results qualitatively agree with experimental data.

\section*{Acknowledgements}
This work has been partially supported by the grant no. PBZ-MIN-008/P03/2003. 


\end{document}